\documentclass[11pt]{article}

\usepackage[totalwidth=460truept,totalheight=620truept]{geometry}
\usepackage{amsfonts,latexsym,amssymb,amsmath,graphicx,accents,slashed,subfigure}
\usepackage[T1]{fontenc}
\usepackage[hidelinks]{hyperref}

\global\arraycolsep=1truept

\begin{document}

\bigskip \phantom{C}

\vskip 1.5truecm

\begin{center}
{\huge \textbf{On The Behavior Of Gravitational Force}}

\vskip.4truecm

{\huge \textbf{At Small Scales}}

\vskip1truecm

\textsl{Marco Piva}

\vskip .3truecm

\textit{INFN, Sezione di Pisa,}

\textit{Largo B. Pontecorvo 3, 56127 Pisa, Italy.}

mpiva@pi.infn.it

\vskip1truecm

\textbf{Abstract}
\end{center}

We point out the idea that, at small scales, gravity can be described by the standard degrees of freedom of general relativity, plus a scalar particle and a degree of freedom of a new type: the fakeon. 
This possibility leads to fundamental implications in understanding gravitational force at quantum level as well as phenomenological consequences in the corresponding classical theory. 

\vspace{0,3cm}
\hfill\break

\noindent
\emph{This essay received an Honorable Mention in the 2019 Gravity Research Foundation Essays on Gravitation Competition.}

\vfill\eject

Recently, it has been shown that quantum field theory (QFT) can describe degrees of freedom of a new type, called fakeons \cite{LWgrav,UVQG,micro}. The fakeon idea finds its roots in a new prescription for the computation of scattering amplitudes \cite{LWformulation,LWunitarity,fakeons}, inspired by the Lee-Wick models \cite{leewick}, and it is the key ingredient for the formulation of a consistent QFT of the gravitational interactions. One of the main predictions of the new theory is that at very small distances cause and effect, past and future loose meaning. This kind of ``indeterminacy'' is related to the presence of a fakeon and the time scale of its occurency has been estimated to be $\sim 10^{-20}s$ \cite{micro}. Moreover, there is the possibility that the fakeon might affect physics at macroscopic scales. In fact, some of its effects survive in the classical limit and could lead to testable predictions in cosmology and black hole physics. In this essay, we present the main features of the theory of particles and fakeons and we explain how this idea can lead to new insights about the fundamental nature of the gravitational force.

The fakeon idea can be used to give sense to higher-derivative theories. In the case of gravity, it is possible to obtain an essentially unique theory by demanding three fundamental properties: locality, renormalizability and unitarity. Together with symmetries requirements (such as gauge invariance, Lorentz invariance and general covariance), these principles are the same used to build the standard model. The gravity theory obtained from these principles is described by the classical action
\begin{equation}\label{stelle}
S(g)=-\frac{1}{2\kappa^2}\int\mathrm{d}^4x\sqrt{-g}\left[\zeta R+2\Lambda+\alpha \left(R_{\mu\nu}R^{\mu\nu}-\frac{1}{3}R^2\right)-\frac{\xi}{6}R^2\right],
\end{equation}
which is then quantized using the new tecnique. This action is well known to be perturbatively renormalizable, as shown by Stelle \cite{stelle}. However, the fakeon procedure makes the theory physically different from the Stelle one, which is quantized with the standard prescription, leading to a violation of unitarity.

The fakeon prescription is encoded in a simple formula for the computation of the physical amplitudes. Consider an amplitude $\mathcal{A}(p)$ obtained from a Feynman diagram associated to some process at a given loop order\footnote{For semplicity, we restrict to the case of one independent external four mometum $p$. Everything applies to the general case \cite{fakeons}.}. Fixing the value of the space momentum $\bf{p}$, the amplitude can be viewed as a function of the complex energy $p^0$ and may have branch cuts. If those branch cuts are present for real values of $p^0$, which is the physical region, then the amplitude is ill defined there. The usual treatment of those singularities is to approach them from the Euclidean region, namely the imaginary axis, by means of the Wick rotation. This procedure is nothing but the analytic continuation of the Euclidean amplitude from above the branch cut. We deonote it by $\mathcal{A}_+(p)$. At the diagrammatic level, this is obained by defining the propagator with the Feynman prescription, which in the case of a massive scalar field reads
\begin{equation}
\frac{1}{k^2-m^2+i\epsilon}.
\end{equation}   
The fakeon prescription, instead, defines the amplitude in the physical region by means of the so called ``average continuation'' \cite{LWformulation,fakeons}
\begin{equation}\label{AC}
\mathcal{A}_\mathrm{F}(p)=\frac{1}{2}\left[\mathcal{A}_+(p)+\mathcal{A}_-(p)\right],
\end{equation}
where $\mathcal{A}_-(p)$ is the analytic continuation from below the cut. This procedure is the arithmetic average of the two possible analytic continuations. In a Feynman diagram, this is achieved \cite{LWgrav} by prescribing the propagator in the following way
\begin{equation}
\frac{k^2-m^2}{(k^2-m^2)^2+M^4},
\end{equation}
plus some special deformations of the integration domains on the loop momenta in Feynman integrals. Here $M$ is a ficticious scale, pretty much like the $i\epsilon$, to be sent to zero at the end of the computations. The reader is referred to \cite{LWformulation,fakeons} for details.

In ordinary non-higher-derivative theories the two prescriptions are compatible with unitarity \cite{LWunitarity,fakeons}, while if higher-derivative terms are included, the Feynman one is not enough to have a unitary QFT. They also have different phenomenological consequences. Typically, branch cuts of amplitudes are associated to physical processes, like particle productions or particle decays, and the branch points represent the physical thresholds of those processes. The fakeon prescription is real, therefore the imaginary part of the associated amplitude is zero, preventing the particles to be physically produced. However, the real part is non zero and (at one loop) is the same we would obtain by using the Feynman prescription. This implies that the fakeon degrees of freedom can still contribute to the divergent parts of the loop integrals, thus having an impact on the renormalization.

To summarize, the general rule can be phrased as follows. Given a quantum field theory, for each degree of freedom it is possible to choose between two quantization procedures: the Feynman one and the fakeon one. The former allows the degrees of freedom to appear as asymptotic states, the latter constraints them to be only virtual, intermediate states circulating inside the loops.

The peculiarity of the fakeon prescription can be used to turn the degrees of freedom that are responsible for the violation of unitarty in the Stelle theory (also known as ``ghosts'') into fakeons. Therefore, the action (\ref{stelle}), quantized with the new prescription, is both renormalizable and unitary. It propagates the two helicities of the graviton, a massive scalar and a massive spin-2 fakeon.

As anticipated, this theory predicts an indeterminacy of causality at small scales. From standard techniques of QFT it is possible to compute the width of the fakeon. The absolute value of its reciprocal is interpreted as the time interval under which is not possbile to define causality \cite{micro}. Taking the fakeon mass $\sim 10^{11}$GeV, for
definiteness, and including the matter content of the standard model, we obtain a time interval of the order of $4\cdot 10^{-20}$s, in the center-of-mass frame. The magnitude decreases by three orders if we increase the mass by one order. For example, for a mass $\sim 10^{12}$GeV we get $4\cdot 10^{-23}$s. The fakeon mass could be much smaller than the Planck mass, without compromising the evidence we have gathered so far on causality. The amount of energy required by to test the violation of causality is too large, at the moment. On the other hand, some effects of the quantization survive the classical limit, leading to corrections to the classical equations of motion. What happens is somewhat similar to the case of the Abraham-Lorentz force in classical electrodynamics \cite{jackson}. There, the equation of motion
\begin{equation}
\left(1-\tau\frac{\mathrm{d}}{\mathrm{d}t}\right)ma=F_{\mathrm{ext}}
\end{equation}
is turned into
\begin{equation}
ma=\frac{1}{\tau}\int_t^{\infty}\mathrm{d}t^{\prime}e^{(t-t^{\prime})/\tau}F_{\mathrm{ext}}(t^{\prime})\equiv\left<F_{\mathrm{ext}}\right>,
\end{equation}
where the external force is ``averaged'' by means of its convolution with the Green's function of $\left(1-\tau\mathrm{d}/\mathrm{d}t\right)$, which is fixed to have a regular limit for $\tau\rightarrow 0^+$. The outcome is a Newton's law with an ``effective force'' $\left<F_{\mathrm{ext}}\right>$.
In the case of fakeons, an analogous procedure applies. The difference is that the prescription which leads to the averaged external force comes from the consistency of the quantum theory, while for the Abraham-Lorentz force it is obtained just by demanding the regularity of $\left<F_{\mathrm{ext}}\right>$ for $\tau\rightarrow 0^+$. In the classical limit, the average continuation (\ref{AC}) reduces to the half sum of the advanced and retarded Green's functions. Therefore, in gravity, the consistency of the quantum theory predicts that the Einstein's equations are modified by an ``effective energy-momentum tensor''
\begin{equation}
R_{\mu\nu}-\frac{1}{2}g_{\mu\nu}R=8\pi G\left<T_{\mu\nu}\right>_{\mathrm{F}},
\end{equation}
where the subscript F reminds that the average is obtained from the fakeon prescription.
This opens new possibilities to test the predictions of this approach \cite{classlim}. One option could be to study the solutions of the modified equations and compare them with those of general relativity \cite{flrw}. Then, it is interesting to investigate situations where the new quantum effects get amplified and become detectable. Moreover, the action (\ref{stelle}) also includes the Starobinsky inflationary model \cite{staro}. In fact, while the fakeon has the role to keep gravity renormalizable, the scalar associated to the $R^2$ term is also responsible for inflation. Very likely, their interaction could bring to new effects in cosmology. Recent results from PLANCK 2018 \cite{planck}, as well as prospective new data in the next future, could be the quickest way to test the fakeon model. In the era of multimessenger astrophysics and new progress in cosmological experiments \cite{cosmoexp}, this might be the best opportunity to investigate new predictions beyond scattering processes and maybe beyond the perturbative level. 

\vfill\eject
\noindent {\large \textbf{Acknowledgments}}

\vskip 2truept

We are grateful to D. Anselmi for reading the manuscript and E. Bianchi for useful discussions. We also thank the Institute for Gravitation \& the Cosmos (Penn State University) for hospitality and the foundation ``\emph{Angelo Della Riccia}'' for financial support.

\end{document}